\documentclass{moriond}

\def\be{\begin{equation}}
\def\ee{\end{equation}}
\def\bea{\begin{eqnarray}}
\def\eea{\end{eqnarray}}

\usepackage{amssymb}
\usepackage{hyperref}
\newcommand{\person}[1]{\textcolor{magenta}{#1}}

\newcommand{\Photo}{}

\begin{document}
\vspace*{4cm}
\title{Theoretical Summary -- Moriond QCD \& High-Energy Interactions 2025}


\author{ P. Skands}

\address{School of Physics and Astronomy, Monash University,
  Wellington Road,\\
Clayton VIC-3800, Australia}

\maketitle\abstracts{The theory talks at Moriond QCD and High-Energy
  Interactions 2025 covered the full
  range of scales from BSM, top, Higgs, EW, and hard QCD physics, through
  resummation, factorisation, and PDFs, 
  to hadronic, heavy-ion, nonperturbative, and lattice QCD. A few
  talks also touched on methodologies. We here summarise main points
  of most of these contributions.}

\section*{Overview}

This summary is divided into the following sections:

\begin{enumerate}\itemsep0.1em 
\item Hard Processes
\item Resummation, Factorisation, and PDFs
\item Quark Flavour Physics
\item Strong Coupling
\item Nonperturbative and Lattice QCD
\item Heavy-Ion Physics
\item BSM Physics
\item Methodology
\end{enumerate}
Some of the talks that involved both experimental and theoretical
aspects may feature only in the experimental summary, by
\person{L.~Shchutska}. For talks that could have been classified
under multiple of the headings above, a choice has been made according
to best judgement. 

\section{Hard Processes}\label{sec:hardproc}

\person{S.~Jaskiewicz}~discussed the large difference between 
di-Higgs cross sections computed using the on-shell (OS)
and $\overline{\mbox{MS}}$ schemes for $m_\mathrm{top}$. In the
high-energy limit, a careful analysis of NLO amplitude structures
reveals that the main difference boils down to
running-$m_\mathrm{top}$ effects, the log structure of which is known
and universal. Pointing out that the truncation of these logs at a
fixed order in the OS scheme is artificial and performing an EFT-like
restoration of them\cite{Jaskiewicz:2024xkd} substantially reduces
the scheme dependence. This seems promising  for future studies. 

\person{M.~Kerner}~pointed out that the differential cross section of
$gg\to H$ vs.\ $p_{T,H}$ allows to distinguish between BSM and
top-quark loop contributions (especially at high $p_{T,H}$), and
demonstrated this in an NLO study of $H+\mathrm{jet}$ with full $m_t$
dependence\cite{CampilloAveleira:2024rnp}.  

\person{M.~Marcoli}\ presented a calculation of $pp\to
\gamma\gamma+\mathrm{jet}$ at NNLO\cite{Buccioni:2025bkl}, using
antenna subtraction within the NNLOJET
framework\cite{NNLOJET:2025rno}, and including $gg$ loop contributions  
at (N)LO. A thorny issue there is the difference between experimental
and theoretical definitions of photon isolation. The new calculation
does yield improved agreement with data but a systematic
undershooting remains, leading to questions of incorporating a 
realistic fragmentation contribution, and/or going to N3LO. The former
would appear worth trying first as it would address the
actual known issue and might involve developing appropriate matching
methodologies which could be of broad use. 

For remarks on the talk of \person{D.~d'Enterria}\ on rare $H$ decays,
see the experimental summary by \person{L.~Shchutska}.

\section{Resummation, Factorisation, PDFs}\label{sec:resPDFs}

\person{S.~Ferrario Ravasio} reported on the completion of the first
NNLL accurate parton shower\cite{FerrarioRavasio:2025soo}, implemented
in PANSCALES~0.3\cite{vanBeekveld:2023ivn}, and 
discussed its building blocks in terms of: 1) NLO matching of the Born
process,  2) correct $\alpha_s^2$ rates of neighbouring pairs of
soft-collinear emissions, soft large-angle emissions, and single collinear
emissions, and 3) the correct $\alpha_s^3$ rate for single
soft-collinear emissions. First results for $e^+e^-$ event shapes show
marked improvements when going from NLL to NNLL, which seems likely to
have beneficial follow-on effects also for improving MC fragmentation
tunes at this accuracy level.

\person{N.~Schalch} discussed the challenges of resumming non-global
logs, exemplified by the problem of calculating the fraction of events
with transverse energy $E_T < Q_0$ in a rapidity gap. His
approach\cite{Becher:2023vrh,Schalch:2024ylx} was to solve the
complicated RGE equation using Monte Carlo methods, resulting in a
shower-style evolution algorithm, dubbed MARZILI. By  
careful insertions of $\Gamma^{(2)}$ (RR $\oplus$ RV $\oplus$ VV),
he achieved NLL accuracy, validated at the 1\% level against
GNOLE\cite{Banfi:2021xzn} 
and PANSCALES\cite{vanBeekveld:2023ivn}. Next steps are a gitlab
release and LHC phenomenology.  

\person{T.~Becher} presented the first resummation of super-leading
logarithms (SLLs)\cite{Becher:2023mtx,Becher:2024nqc}, emphasising that
one needs full-colour results to see the effect. The SLLs 
were computed analytically, order by order, then summed. As expected
the biggest contributions come from gluonic 
channels, and reach $10\%$ for small values of the gap-fraction 
$Q_0$ value. There now remains to combine these with the
non-global logs presented by \person{Schalch} (at full colour) before
comparing to LHC measurements. 

\begin{figure}[t]
  \centering
  \includegraphics*[scale=0.5]{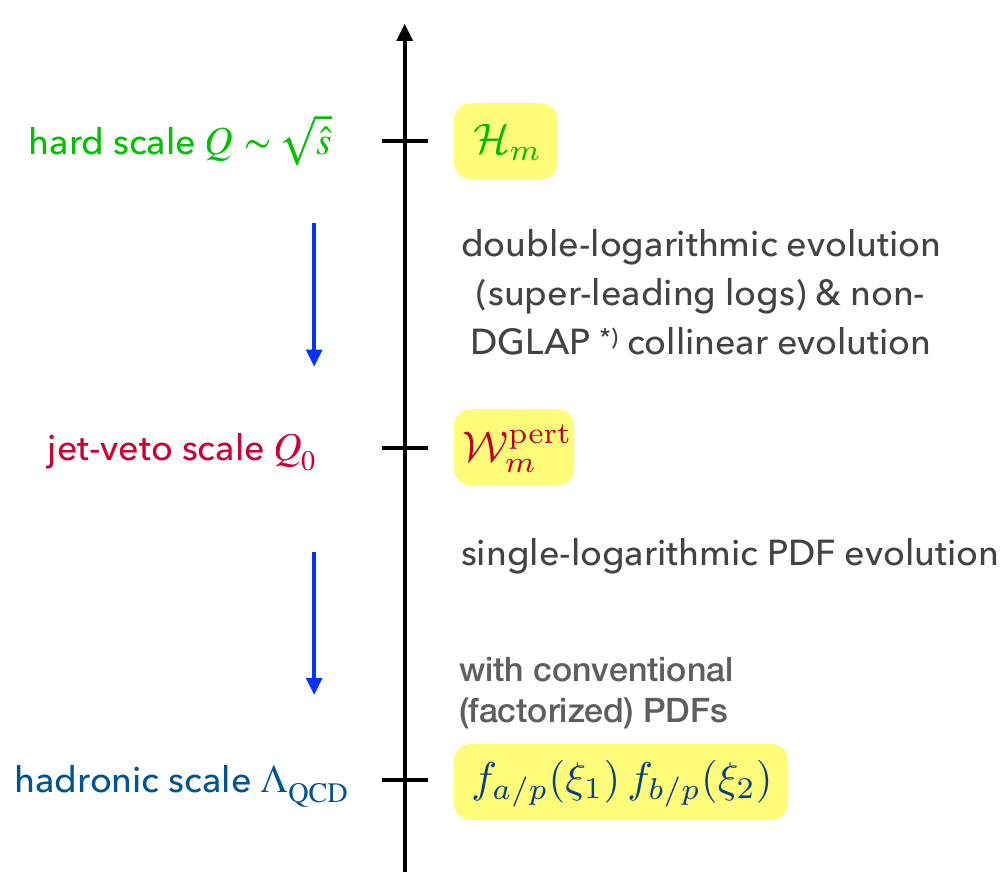}
  \caption{Illustration (adapted from the talk by \person{M.~Neubert})
    of factorisation restoration in gaps-between-jets.
  \label{fig:neubert}}
\end{figure}

\person{M.~Neubert} turned the longstanding question of factorisation
violation into one of factorisation restoration\cite{Becher:2024kmk}.
Specifically, the collinear factorisation that e.g.\ PDF factorisation
relies on is known to be broken for spacelike (i.e., initial-state)
splittings at 3-loop order.
Through careful analysis of a gaps-between-jets
setup, the factorisation-violating contributions were shown to be
associated only with scales greater than the veto scale,
with conventional collinear factorisation restored for scales $\mu <
Q_0$. This amounts to an explicit proof at 3-loop order
that factorisation works for scales below $Q_0$, above which 
super-leading and non-DGLAP contributions must be taken into
account; see the illustration in Figure~\ref{fig:neubert}. 

\person{A.~Pelloni} reported on the last missing piece required for
N3LO PDF evolution: an OPE-based approach from which
the first ten Mellin moments ($N\le 20$) of the 4-loop $g\to gg$
splitting function have been determined, along with partial
information on the higher moments\cite{Falcioni:2024qpd}.
The resulting uncertainty from scale variations and the ambiguity on the higher
moments remains below 1\% for $x > 10^{-4}$ (except near $x\sim 0.7$
where a sign change occurs), enabling precise PDF evolution at N3LO.   

\person{F.~Hekhorn} noted that, from the practical point of view of
PDF fits, N3LO evolution can be regarded as
completed\cite{MSHT:2024tdn,Hekhorn:2025xke} (though N3LO 
\emph{cross sections} are another story). He emphasised
that QED effects tend to reduce Higgs cross sections by $\sim 1\%$
and that this behaviour is consistent across MSHT and NNPDF fits.

Within the framework of transverse-momentum-dependent (TMD) PDFs,
\person{A.~Vladimirov} presented an all-orders resummation of 
kinematic power corrections (KPCs) $\propto (k_T/Q)^n$ for the angular
coefficients of Drell-Yan\cite{Piloneta:2024aac,Vladimirov:2025qys},
in good agreement with measurements, while 
\person{W.~Zhan} used a Monte-Carlo parton-branching method to evolve
TMD PDFs, with a Gaussian assumption for the ``intrinsic''
transverse-momentum component $k_T/q_s$ only affecting the 
low-$p_T$ part of the spectrum; $q_s$ could then be extracted from the
ratio between low and high $p_T$\cite{Zhan:2025wup},
resulting in $q_s \sim 1\,$GeV for low $\sqrt{s} < 10\,$GeV,
increasing for larger 
$\sqrt{s}$. We note that similar conclusions are obtained with
full-fledged shower Monte Carlos\cite{Skands:2014pea}. 

\section{Quark Flavour Physics}\label{sec:qFlav}

\person{N.~Gubernari} pointed out that the standard (BGL) parameterization of
form factors for $B$ decays yields a divergent series whenever there
are on-shell channels \emph{below} the one in question (e.g., for
$B\to K$ FFs, such a ``subthreshold'' channel is $B\to \pi$). This renders the
truncation error meaningless. He then noted that the analytic
structure suggests an alternative (GG) parameterization, which
produces a reliable truncation error\cite{Gopal:2024mgb}.

\person{D.~Mishra} considered the issue of the measured
$\mathrm{BR}(B\to K\mu\mu)$ being lower than theory predictions (which
also exhibit tensions amongst
themselves). From a light-cone sum-rule analysis of the $B\to K$
transition ME, one conclusion was that non-factorisable soft-gluon
contributions to the charm loop can safely be neglected. The tension
with the measured result persists in their re-analysis of the hadronic
ME\cite{Mishra:2025tia}.  

\person{E.~Lunghi} highlighted that the exclusive $b\to s\mu\mu$ modes
are subject to potentially large and uncontrolled power corrections
while the inclusive rate is free of hadronic uncertainties up to 
${\cal O}(\Lambda_{\mathrm{QCD}^2}/m_b^2)$. Although the OPE does
break down at large $q^2 = m_{\mu\mu}^2$\cite{Huber:2023qse}, it still
holds for the ratio 
of (integrals of) $B\to X_s\ell \ell$ to $B\to X_u\ell \nu$. LHCb
already has enough data to produce a corresponding measurement 
for $q^2 > 14.4\,$GeV$^2$. Meanwhile, a preliminary ``frankenfit''
indicates that the inclusive modes are currently in agreement with
data\cite{Huber:2024rbw}. 

\person{S.~Nandi} pointed out that tests of lepton flavour
universality analogous to $R_{D^{(*)}}$ can be done with
$B_c \to \eta_c,~J/\psi,~\chi_c,~\ldots$. For $B_c \to J/\psi$, there exist FFs
from lattice, which can be converted to ones for $B_c\to\eta_c$ using
heavy quark spin symmetry, with an estimated symmetry-breaking
correction of $\sim 30\%$. $B_c \to$ P- and S-wave FFs were also
computed in both NRQCD and pQCD, valid for low and high $q^2$
respectively\cite{Dey:2025xdx}.

Moving from $B$ mesons to $K$ ones, \person{G.~d'Ambrosio} presented
an analysis of $K\to \pi\ell\ell$, noting that it is dominated by long
distances but that, at leading colour, it simplifies to a VMD-like sum of
1-meson poles which converges to a $\log(q^2)$ behaviour for large
$q$\cite{DAmbrosio:2024ncc}. 

\section{Strong Coupling}\label{sec:alphaS}

\person{M.~Benitez} reported on a re-analysis of the
extraction of $\alpha_s$ from the heavy jet
mass\cite{Benitez:2025vsp,Benitez:2025udh}, which historically 
had yielded low values and was known to be affected by significant
hadronization corrections. With a new and sophisticated treatment of
experimental and theoretical correlations and a thorough treatment of
the impact of the fit range, the result, albeit still slightly low at
$\alpha_s(m_Z) = 0.1145^{+0.0021}_{-0.0019}$, is compatible with those
from other event shapes. 

\person{J.~Pires} presented an extraction from dijets at LHC and
HERA, pointing out that one gets sensitivity not only from the cross
sections but also via the PDFs. Using NNLOJET\cite{NNLOJET:2025rno}
with reduced scale 
dependence and subleading-colour contributions for the first time in
an LHC $\alpha_s$ measurement, with central scales $\mu^2_{\mathrm{LHC}}
= m^2_{jj}$ and  $\mu^2_{\mathrm{HERA}} = Q^2 +
\left<p_T\right>_{1,2}^2$, the combined result is $\alpha_s(M_Z) =
0.1180 \pm 0.0010_\mathrm{fit,PDF}\pm 0.0001_{\mu_0}\pm
0.0022_{\mu_R,\mu_F}$\cite{Ahmadova:2024emn}. 

\person{P.~Petreczky} reviewed lattice determinations of $\alpha_S$
from FLAG 2024\cite{FlavourLatticeAveragingGroupFLAG:2024oxs},
emphasising the pros and cons of three dominant 
methodologies: 1) determination using physical quantities for which
the requirement of adequate resolution for a fairly large lattice
size results in the ``window problem'', $\mu a \ll 1, L = N_s a >
1\,\mathrm{fm} \Rightarrow \mu = 1 - 3\,$GeV, resulting in somewhat
large systematic errors; 2) comparison with lattice perturbation
theory $\Rightarrow \alpha_s(\mu = 1/a)$ limited by the
accuracy of lattice perturbation theory; 3) lattice QCD in ``femto
boxes'' and special schemes, $\mu \sim 70\,$GeV, $\mu a \ll 1$, $N_s a
\ll 1\,$fm, connected to physical calculations by a procedure called
``step scaling''. The FLAG 2024 result is $\alpha_s(M_Z) =
0.1183(7)$. When including the PDG average this changes only
slightly, to $\alpha_s(M_Z) = 0.1181(7)$. 

\person{Y.~Che} and \person{J.~Wu} presented $\alpha_s$
determinations from inclusive semileptonic $B$ and $D$ decays
respectively\cite{Che:2024kad,Wu:2024jyf}. In both cases, a 
a heavy-quark expansion was used, with $|V_{cb}|$ and $|V_{cs}|$
respectively then having to be 
taken from other measurements. For the former, it seems possible to reach
a competitive uncertainty of $\Delta \alpha_s(M_Z) \sim 0.0018$. For
the latter, noting that the semileptonic partial widths for $D^0$ and $D^+$ are
the same within the overall uncertainty of a few percent, these may be
combined yielding a first determination of $\alpha_s$ at a scale below
$m_\tau$, of $\alpha_s(m_c) = 0.445 \pm 0.009_{\mathrm{exp}} \pm 0.114$, while the
value extracted from $D_s$ decays is $\alpha_s(m_c) = 0.400 \pm 0.014_\mathrm{exp}
\pm 0.113$. Both values agree with the corresponding world average
within uncertainties.  

\section{Nonperturbative and Lattice QCD}\label{sec:NPLatt}

\person{B.~Toth} presented the latest update on $g_\mu-2$, noting that
the largest theoretical uncertainty comes from the hadronic vacuum
polarisation (HVP) which has traditionally been determined in two main
ways, either data-driven from the $R$-ratio in $e^+e^- \to
\mathrm{hadrons}$ via the optical theorem, or from lattice QCD
calculations. For the former, there have been some tensions among
different data sets. The theory reference from the 2020 so-called
white paper\cite{Aoyama:2020ynm} used measurements by BaBar and
KLOE. More recent determinations from Tau decays and from 
CMD-3, not used in the white paper, translate to values of $g_\mu$
significantly closer to the BNL+FNAL results. Meanwhile, the lattice
community have reached an agreement on an intermediate (simpler) benchmark
`'window observable'' to cross check calculations, and have also moved
to finer lattice spacings. The most recent update is a combined fit
using lattice for an 0-2.8 fm window and the data-driven method for
distances larger than 2.8 fm (which amounts to 5\% of the total
result, avoids the $\rho$ peak, and exhibits good agreement between
different measurements). This moves the central BMW/DMZ 2024 theory
prediction\cite{Boccaletti:2024guq} even closer to the BNL+FNAL
average, translating to a current difference of $0.9\sigma$. The
conclusion is that there is no indication of new physics in $g_\mu - 2$.

\person{S.~Zafeiropoulos} explained the breakthrough\cite{Ji:2013dva}
that has allowed 
lattice to provide ab-initio PDF determinations without theoretical
obstructions\cite{Joo:2020spy,HadStruc:2021wmh}, essentially by putting partons
some distance apart, boosting them to achieve almost lightlike
separation (in the proton frame) and then using a perturbative mapping
from finite to infinite momentum\cite{Ji:2013dva}. The method is also
able to provide helicity and transversity PDFs, as well as PDFs for
other hadron species such as pions. 

\person{S.~Li} took note of the experimental observation that there are more
low-$z$ $D^*$ mesons in data than in the baseline MCs, and
investigated whether this could originate from recombination of $c$
quarks with quarks from the underlying
event\cite{Jiang:2023pvj}. Simple assumptions for 
this ``UE sea'' yielded good fits to data, prompting the question
whether a similar effect involving recombination with diquarks could
account for observed enhancements of low-$p_T$ heavy-flavour
baryons. We note that similar phenomena are investigated in the
contexts of coalescence and QCD colour reconnections\cite{Altmann:2024kwx}.

\person{M.~Praszalowicz} observed that the ``dip-bump'' structure of
the differential cross section for elastic scattering appears to be
universal between ISR and LHC energies, and derived a corresponding
formula for the scaling of the elastic $\rho$
parameter\cite{Baldenegro:2024vgg,Praszalowicz:2025vdb}.  

\person{R.~Sandapen} considered whether the pion could be understood
from the point of view of
holography\cite{Forshaw:2024mrh}. Conventionally, this picture 
relies on light-front quantisation with longitudinal degrees
of freedom being neglected, but that results in a massless pion. By
restoring the conformal-symmetry breaking longitudinal potential
$U_{||}$ one can obtain not only the correct $m_\pi$ but also $f_\pi$,
$r_\pi$, and low-$Q$ form factors. Three different choices for
$U_{||}$ lead to slightly different values for
$\Gamma_{\gamma\gamma}$, with the one closest to the experimental
value also exhibiting quantitative agreement with the holographic
prediction in the limit of weak coupling.

\section{Heavy Ions}\label{sec:HI}

The heavy-ion session was introduced by \person{N.~Zardoshti}, who
gave a comprehensive standalone summary of the field.

\person{I.~Kolb\'e} presented an elaborate simulation of jets in
medium\cite{Kolbe:2023rsq}. Noting that the observation of nonzero
$v_2$ at high $p_T$ is 
not well understood, she set out to answer if this could be due to
misestimated hydro effects in Monte Carlo generators such as
JEWEL\cite{Zapp:2013vla}, which 
simulates in-medium jet evolution. Extending JEWEL with a careful hydro
description delivered by TRAJECTUM\cite{Nijs:2020roc}, the conclusion
is that this agrees 
well with the jet $p_T$ spectrum in central Pb-Pb collisions, but that
the non-zero $v_2$ at high $p_T$ is still not well understood. 

\person{E.~Speranza} discussed that the huge initial-state angular
momentum in peripheral heavy-ion collisions may be converted into
polarisation\cite{Giacalone:2025bgm}. This calls for relativistic spin
hydrodynamics with 
conservation of angular momentum. Conclusions were that 
initial-state spin fluctuations can be much larger than those induced
by final-state vorticity, and that a useful experimental probe is
$\Lambda\Lambda$ correlations.

Topics in heavy-ion physics were also touched on in the joint
session with gravity, in which \person{W.~Schee} discussed interplays
between QCD and gravity. 

\section{Beyond the Standard Model}\label{sec:BSM}

\person{S.~Vempati}~presented an update on the constraints on FLV Higgs couplings to
charged leptons\cite{Abu-Ajamieh:2025jsz}. In particular for 
$|Y_{\tau \mu}|$, the vast numbers of Higgs bosons produced at the LHC
already yield $|Y_{\tau \mu}| \lesssim {\cal O}(10^{-3})$ which
surpasses the constraints from lower-energy probes such as $\tau \to
\mu\gamma$,  $\tau \to 3\mu$ and EDM$_\mu$. The HL-LHC is expected to
yield a further order-of-magnitude improvement.

\person{E.~Vryonidou}~reported on a comprehensive study of CPV effects in the
Higgs sector within a SMEFT framework. Main points included the 
complementarity of $t\bar{t}H$ and $tHq$ processes for determining the
top Yukawa coupling\cite{Miralles:2024huv}, and the importance of
angular observables for 
determining CP violating gauge-Higgs couplings in $VH$ (at
NLO)\cite{Rossia:2024rfo}, and loop-induced $gg\to hh$, $hZ$, $WW$,
and $ZZ$ processes\cite{Thomas:2024dwd}.  

Three further talks in the area of BSM were presented, by
\person{M.~Baker}\ (on
heavy vectors\cite{Baker:2024xwh}), \person{S.~Balan}\ (on
  dark-matter fits\cite{Balan:2025uke}), and by \person{M.~Fedele}\ (on 
sterile neutrinos in $b\to c\ell^+\ell^-$\cite{Bernlochner:2024xiz}). 

\section{Methodology}\label{sec:meth}

\person{R.~Poncelet} remarked on the importance of improving the
reliability and comprehensiveness of perturbative uncertainty estimates,
and presented a proposal cast in terms of a generic expansion  
with a priori unknown ``theory nuisance
parameters''\cite{Lim:2024nsk}. The most complex process shown 
was $pp\to ZZ^* \to ee\mu\mu$, for which the NLO and NNLO
uncertainties did exhibit improved consistency, though the
failure at LO was unchanged. 

\person{M.~White} proposed to employ a measure of quantum advantage,
called ``magic'', to the study of spin correlations in $t\bar{t}$
events\cite{Aoude:2025jzc}. 

\section*{Acknowledgements}
PS is supported by the Australian Research Council, grant DP220103512.

\section*{References}
\bibliography{skands}


\end{document}